\documentclass[a4paper,sort,superscriptaddress,preprint]{revtex4}

\usepackage{graphicx}
\usepackage{amsmath}
\usepackage{amsfonts}
\usepackage{mathrsfs}
\usepackage{bbold}

\begin{document}

\title{Quaternions and hybrid nematic disclinations}

\author{Simon \v{C}opar}

\affiliation{
Faculty of Mathematics and Physics, University of
  Ljubljana, Jadranska 19, 1000 Ljubljana, Slovenia
}

\affiliation{
Department of Physics and Astronomy, University of Pennsylvania,
  209 South 33rd Street, Philadelphia, Pennsylvania 19104, USA
}

\author{Slobodan \v{Z}umer}
\affiliation{
Faculty of Mathematics and Physics, University of
  Ljubljana, Jadranska 19, 1000 Ljubljana, Slovenia
}
\affiliation{
J. Stefan Institute, Jamova 39, 1000 Ljubljana, Slovenia
}

\date{\today}
\begin{abstract}
Disclination lines in nematic liquid crystals can exist in different
geometric conformations, characterised by their director profile. In
certain confined, colloidal and even more prominently in
chiral nematics, the director profile may vary along the disclination
line. We construct a robust geometric decomposition of director
profile variations in closed disclination loops based on a quaternion
description and use it to apply
topological classification to linked loops with arbitrary variation of
the profile. The description bridges the gap between the known
abstract classification scheme derived from homotopy theory and the
observable local features of disclinations. We compare the resulting
decomposition of disclination loop features to a similar decomposition
of nematic textures on closed surfaces.
\end{abstract}


\newcommand\ii{\mathbb{i}}
\newcommand\jj{\mathbb{j}}
\newcommand\kk{\mathbb{k}}
\renewcommand\vec[1]{\mathbf{#1}}

\maketitle

\section{Introduction}

The defects in condensed materials have a significant effect on their
physical properties. While their existence may be caused by impurities
and flawed preparation process, controlled manipulation of defects can
enable tuning of the physical characteristics of the material. The
ability of a medium to host defects is a side-effect of symmetry
breaking on the microscopic scale. As such, the defects are intimately
related to the topological properties of degenerate microscopic
degrees of freedom.  In materials without broken translational
symmetry, such as uniaxial and biaxial nematics
\cite{degennes_knjiga}, ferromagnets \cite{kotiuga} and superfluid
helium \cite{quant_Helc,volovik_mineev}, the defects can be roughly
classified by their dimension into point defects, line defects and
walls \cite{points_walls_knjiga}. Their classification with the aid of
homotopy groups is a well-established and efficient way of describing
their interactions and the conservation rules they obey
\cite{mermin}. However, the topological analysis does not account for
the fine geometric details, which are under control of the free
energy.

Specifically for nematic braids -- networks of closed nematic
disclination lines
\cite{araki_collagg,ravnik_softmatter_entangled,ravnik_dimers_wires}
-- recent development has shown that additional geometric constraints
allow for a finer classification of nematic braids and simplifies the
understanding of their rewiring
\cite{copar_rewiring,uros_science}. The constraint of a fixed $-1/2$
disclination profile that holds for nematic braids is not universal
and does apply necessarily to some constrained \cite{araki_nmat_mem}
or chiral systems
\cite{junichi_wedge,fukuda_blue_prl,fukuda_blue_ring}. But even
without constraints, it should still be possible to convey the
topological information through the geometric primitives. In this
paper, we demonstrate a formalism that can describe the most general
types of nematic disclinations in terms of the local director behavior
and enumerate them using quaternions. We show that our interpretation
naturally extends the existing formalism for $-1/2$ disclination loops
and reveal a connection between closed disclination loops and
two-dimensional nematic textures.

\section{Overview}

Before introducing any new results, we shall briefly overview the
existing theoretical background regarding nematic defect loops.

A nematic liquid crystal consists of elongated molecules that are
locally orientationally aligned \cite{degennes_knjiga}. In a continuum
approximation, the order is described by the director $\vec{n}$, a
unit vector field, pointing in the direction of average molecular
orientation. The head-to-tail symmetry of the molecules causes the
sign of the director to be ambiguous.

Topologically, the director field is a map from the coordinate space
$\mathbb{R}^3$ to the real projective plane $\mathbb{R}P^2$ -- the
ground state manifold (GSM). The fundamental group
$\pi_1(\mathbb{R}P^2)=\mathbb{Z}_2$ distinguishes disclination lines,
where the director makes a half-turn around the disclination, from
nondefect states, but makes no distinction between different
disclinations \cite{mermin}.

Our specific interest is focused on disclination lines that are closed
into a loop. Disclination loops may be linked by other disclinations
or carry a topological point charge, measured by the second homotopy
group. Closed loops therefore also carry an additional topological
index $\nu \in \mathbb{Z}_4$. An established result shows
\cite{janich,nakanishi,randy_rmp}, that a set of $n$ disclinations
obeys a law in the form
\begin{equation}
  \frac{1}{2}\left(\sum_{i=1}^n \nu_i - 2\sum_{i>j}^n {\rm Lk}_{ij}\right)=q\mod 2,
  \label{eq:janich}
\end{equation}
where $q$ is the topological point charge of the set of disclination
loops, ${\rm Lk}_{ij}$ are linking numbers between loops and the
indices $i$ and $j$ run over all the loops. Odd indices $\nu$
correspond to disclinations that are threaded by other disclinations
an odd number of times while even indices correspond to those that are
threaded by and even number, which also applies to unthreaded
loops. This theory describes all existing topological properties of
disclination loops. However, the index $\nu$ is an abstract attribute
assigned to each of the loops, so the application to the systems that
are found in experiments, simulations and theoretical models requires
a way to relate this index to observable attributes.

If instead of a three-dimensional space, the molecules of the nematic
are restricted to move and orient in a two-dimensional plane, the
system exhibits point defects, which can have any half-integer winding
number, the lowest ones being $-1/2$ and $+1/2$ \cite{mermin}.  These
two-dimensional point defects represent a restricted set of cross
sections the disclination lines in three dimensions can have. The
topological equivalence of all disclination lines in the
three-dimensional case is caused by the ability of the director to
rotate through the third dimension. If the director is forced by the
free energy minimization or geometric constrants to be perpendicular
to the disclination tangent, which is excellently obeyed in
well-researched nematic colloids
\cite{musevic_science,ravnik_dimers_wires,copar_tine_ribbons}, closed
$-1/2$ disclination loops can be described as ribbons that follow the
rotation of the profile. Ribbons can be assigned an invariant called
the self-linking number ${\rm Sl}$ \cite{geom_calugareanu,pohl_sl},
which in the case of $-1/2$ disclinations quantifies how many times
the three-fold disclination profile rotates around the tangent when we
traverse the loop \cite{copar_rewiring}. In a braid of many
disclination loops, each self-linking number is of the form ${\rm
  Sl}=m/3$, where the numerator $m$ is odd if the loop is threaded by
an odd number of other disclinations, and even if the loop is linked
by nothing or by an even number of disclinations. The self-linking
numbers of a set of $n$ disclination loops with a $-1/2$ profile obey
a rule
\begin{equation}
  \frac{3}{2}\left(\sum_{i=1}^n {\rm Sl}_i+ \sum_{i\neq j}^n {\rm Lk}_{ij}\right)+n=q \mod 2,
  \label{eq:sl_law}
\end{equation}
that bears resemblance to the J\"anich's rule (\ref{eq:janich}). The
integers ${\rm Lk}_{ij}$ are the linking numbers between loops and
${\rm Sl}_i$ are self-linking numbers of individual loops. The
comparison of equations (\ref{eq:janich}) and (\ref{eq:sl_law})
suggests a connection between the index $\nu$ and the self-linking
number. In the following section, we confirm and extend this notion to
all disclination loops.

\begin{figure}[t]
  \centering
  \includegraphics[width=\columnwidth]{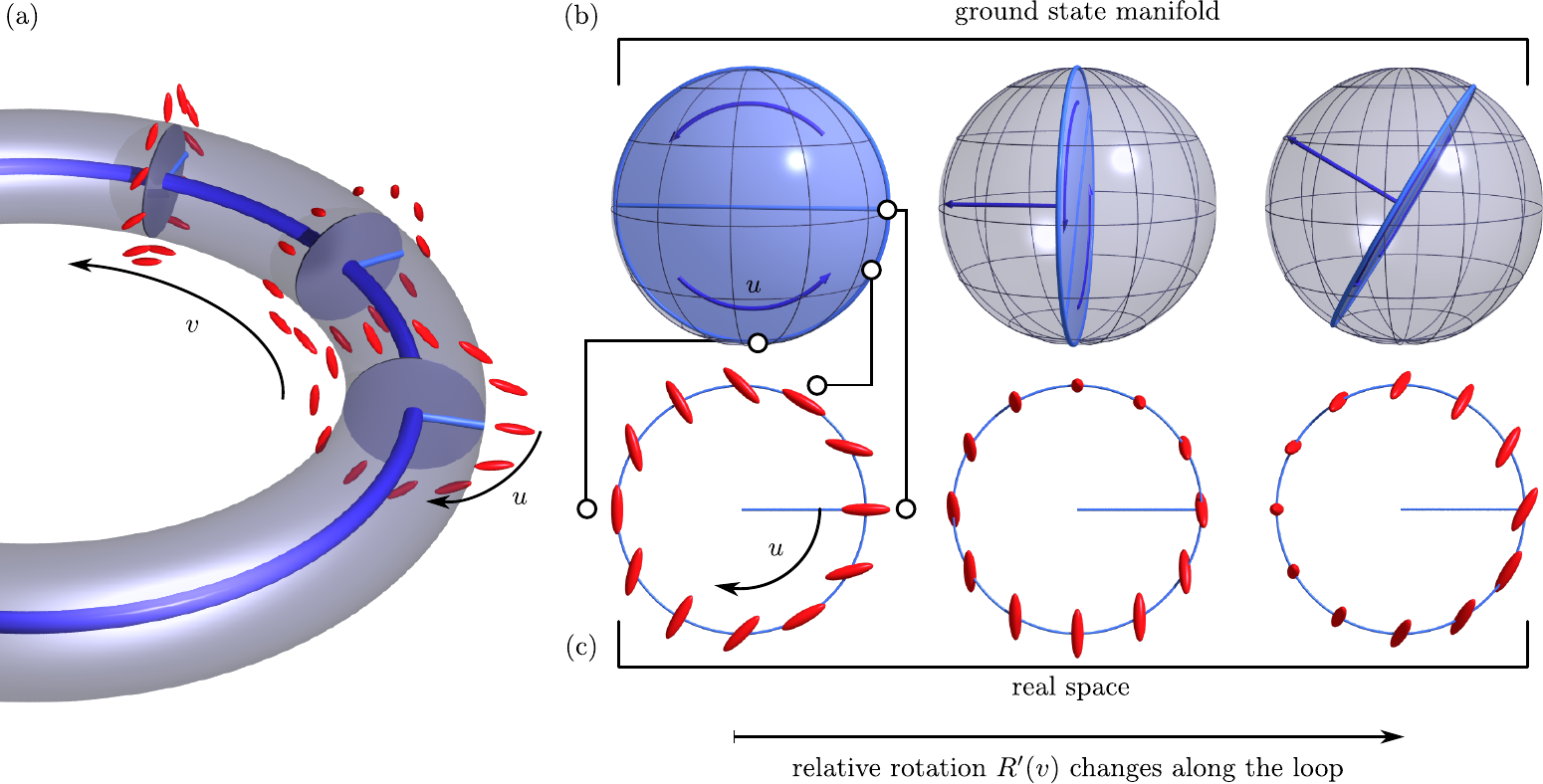}
  \caption{
    \label{fig:figure1}
    (a) The director profile specified on a torus enveloping a general
    disclination loop. The small radius is parametrized by $u$ and
    the large radius is parametrized by $v$, producing a seamless
    doubly periodic parametrization. (b) The director on any small
    cross section lies on a great circle in the GSM. The origin of the
    parameter $u=0$ provides a reference in
    the plane of the circle and the sense of rotation in GSM with increasing
    $u$ orients the circle's normal so that it defines a clockwise
    motion according to the right-hand rule. (c) Each circle in the GSM
    matches a cross section in real space with a fully specified
    parametrization $u$.  (a-c) The orientation of the director cross
    section and its associated circle changes along the disclination with
    parameter $v$, which is described by a continuously changing
    rotation ${\sf R}'(v)$.
  }
\end{figure}

\section{The profile circles and quaternions}

All disclinations in a 3D nematic have in common that the director on
a test loop around the disclination traces an irreducible path in the
ground state manifold (GSM). As the GSM is a sphere with additional
condition of the antipodal points identified, the loop is
topologically equivalent to a half of a great circle, which can be
symmetrically extended to the full circle
(Fig.~\ref{fig:figure1}). As the vicinity of the
disclinations is subject to elastic free-energy restrictions, this is
a quantitatively reasonable approximation.

A closed disclination loop can be wrapped in a torus, $S^1\otimes
S^1$, generated by revolving a disclination-encircling loop
parameterized by $u$, around the disclination loop parameterized by
$v$. For each value of $v$, the parameter $u$ traces a different great
circle in the GSM (Fig.~\ref{fig:figure1}a). Each circle
is fully specified by its normal and a reference director at a chosen
parameter, $u=0$ (Fig.~\ref{fig:figure1}b,c). The sense of
circulation around the circle is important and prescribes the sign of
its normal. For example, the $-1/2$ profile corresponds to the normal
of the circle being anti-parallel to the disclination tangent in the
real space, as the director rotates in the opposite sense compared to
the circulation of the enclosing loop (hence the minus sign in
$-1/2$).  The circle is invariant to rotations for $\pi$ around its
normal because each point in the real space maps to two antipodal
points in the GSM.

Let the parametrization of the director field on the torus be
$\vec{n}(u,v)$. When $v$ changes, the circle rotates, so the director
at all values of parameter $u$ transform with the same rotation,
parameterized by $v$,
\begin{equation}
  \vec{n}(u,v)={\sf R}(v)\vec{n}(u,0).
  \label{eq:gsm_uv}
\end{equation}
The $\vec{n}(u,0)$ is simply a reference profile at a freely chosen
point on the loop. This sets the initial rotation to unity, ${\sf R}(0)=1$.

The rotations vary continuously with loop parameter $v$ and thus
describe a path in the space of all rotations in three-dimensions. We
will use the quaternion representation of the $\mathrm{SU}(2)$ group.
The unit quaternions $\ii$, $\jj$ and $\kk$ obey the conventional
structure relations $\ii^2=\jj^2=\kk^2=-1$, $\ii\jj=-\jj\ii=\kk$,
$\jj\kk=-\kk\jj=\ii$ and $\kk\ii=-\ii\kk=\jj$. A rotation around an
arbitrary axis $\vec{a}$ by an angle $\phi$ follows the usual spinor
formula
\begin{equation}
  {\sf R}=\cos\frac{\phi}{2}+(a_x\ii+a_y\jj+a_z\kk)\sin\frac{\phi}{2}.
  \label{eq:r_spinor}
\end{equation}
Rotation by $2\pi$ around any axis yields ${\sf R}=-1$, which is the
irreducible rotational motion that ends up in the same state where it
started. Note that if we took the $\mathrm{SO}(3)$ representation,
these irreducible paths would be considered equal to the identity
transformation, and some of the topological information would be lost.

The continuity condition for a closed loop is satisfied if the
director is continuous for \emph{all} $u$, that is, if the circle we
obtain after completing the loop at ($v=2\pi$) matches the circle at
the beginning of the loop ($v=0$) up to a $\pi$ rotation around the
loop tangent, which we align with the $z$-axis. Because the
$\mathrm{SU}(2)$ group covers the space of rotations twice, the set of
possible cumulative rotations expands to
\begin{equation}
  {\sf R}(v=2\pi)\in\{1,\kk,-1,-\kk\}=\kk^\nu.
  \label{eq:quatern_cycle}
\end{equation}
The allowed rotations form a cyclic group $\mathbb{Z}_4$, enumerated
by an integer $\nu$, which can be shown to correspond to the index
introduced by J\"anich (\ref{eq:janich}). As ${\sf R}(2\pi)$ is simply
the total rotation of the director at any fixed $u$ on a circuit
around the disclination, it measures the fundamental group of the
nematic encircled by the disclination. ${\sf R}=\pm\kk$ are nontrivial
rotations by $\pi$ and mean the disclination is threaded by another
disclination. ${\sf R}=1$ and ${\sf R}=-1$ correspond to an unlinked
loop with even and odd topological charge, respectively (this topic is
discussed thoroughly in \cite{randy_rmp}).

Each rotation can be decomposed into the rotation ${\sf R}_0(v)$ of
the local coordinate frame relative to the global coordinate frame,
and the rotation ${\sf R}'(v)$ of the director relative to the local
coordinate frame.
\begin{equation}
  \vec{n}(u,v)={\sf R}_0(v){\sf R}'(v)\vec{n}(u,0).
\end{equation}
We set the local coordinate frame to keep its $z$-axis aligned to the
disclination loop tangent and vary continuously along the loop
(e.g. the Frenet-Serret frame). We also ensure our framing is
topologically equivalent to a framing of a planar circular loop
without torsion, so the local coordinate frame simply rotates by
$2\pi$ when we traverse the loop: ${\sf R}_0(2\pi)=-1$. This extra
rotational offset is present for all the loops, so we can factor it
out. The multiplication of $\kk^\nu$ by $-1$ simply shifts $\nu$ by
$2$ so an alternative parameter $\nu^\ast=\nu+2\mod 4$ can be
introduced, rewriting the J\"anich's law (\ref{eq:janich}) as
\begin{equation}
  \frac{1}{2}\left(\sum_{i=1}^n \nu_i^\ast - 2\sum_{i>j}^n {\rm Lk}_{ij}\right)+n=q\mod 2.
  \label{eq:janichast}
\end{equation}
We can see that the number of loops $n$ plays a role in the
conservation law because a trivial disclination loop with $\nu^\ast=0$
holds a nonzero topological charge (one example is a Saturn ring
defect). All the significant information is now encoded in ${\sf
  R}'(v)$, which is measured in a convenient disclination-aligned
local coordinate frame.

\section{Algebra of profile transitions}

\begin{figure}
  \centering
  \includegraphics[width=\textwidth]{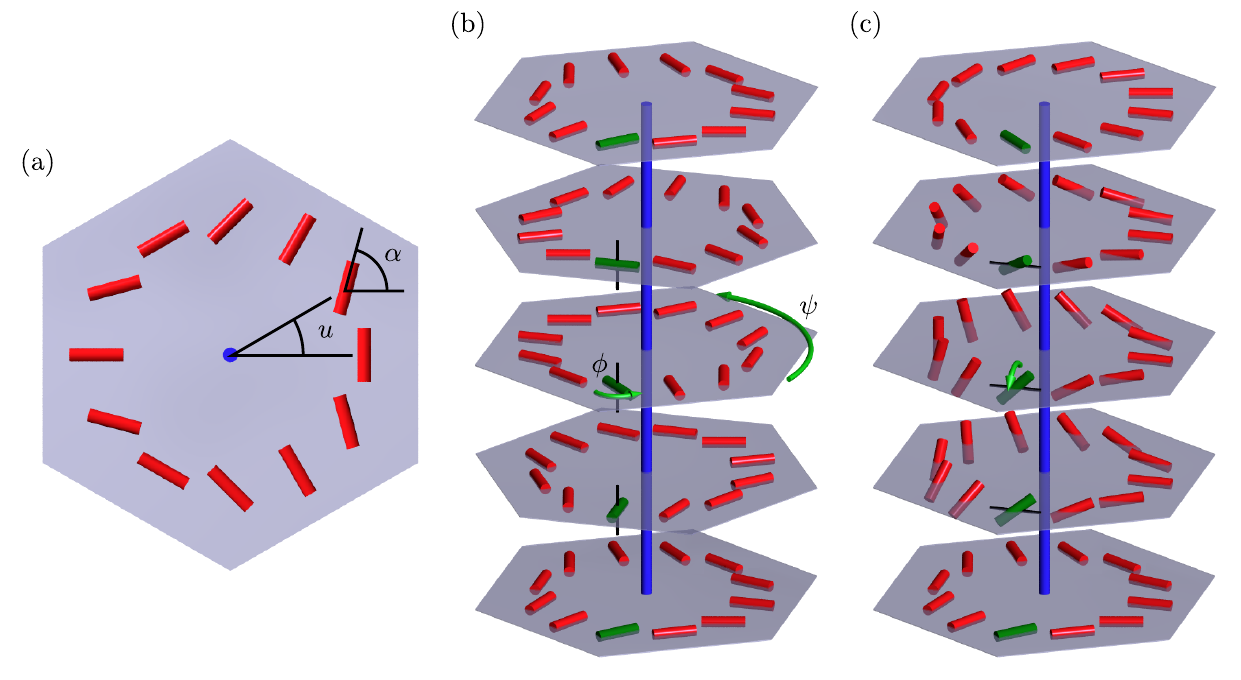}
  \caption{
    \label{fig:figure2}
    (a) Notation used in equation (\ref{eq:alpha}). The $-1/2$ profile is two-dimensional, so the director can be represented
    by a single angle $\alpha$ that varies with parameter $u$.
    (b) Rotation of the director around the $z$ axis induces a rigid rotation of the $-1/2$ disclination profile.
    Depicted is a rotation of the director for $\phi=\pi$, which yields a $1/3$ of a full turn for the profile ($\psi=2\pi/3$).
    (c) The transition between $+1/2$ and $-1/2$ disclination profiles
    is realised by a rotation for $\pi$ around a perpendicular axis ($x$ axis, marked in black).
    There are two senses of rotation -- the one depicted has the parity $P=+1$.
  }
\end{figure}

Each geometric operation on the disclination profile can now be
directly translated into the quaternion language. First, we reproduce
the result for the $-1/2$ disclination profile. In the relative frame
along the curve, the director lies in the $xy$ plane and can only
rotate around the $z$-axis, as it has to stay perpendicular to the
disclination tangent. Rotation of the profile consists of both the
passive rotation of the coordinate frame and the active rotation of
the director. As the director lies in a plane, it can be specified by
an angle $\alpha$, so the profile rotation by $\psi$ yields a
$\phi=3/2\psi$ rotation of the director, and subsequently the entire
circle (Fig.~\ref{fig:figure2}a,b):
\begin{equation}
  \alpha(u)=-\frac{1}{2}u;\quad
  \alpha'(u)=-\frac{1}{2}(u-\psi)+\psi=\alpha(u)+\underbrace{\frac{3}{2}\psi}_{\phi}.
  \label{eq:alpha}
\end{equation}

In a loop with a self-linking number ${\rm Sl}$, the profile rotates
by a total angle of $\psi=2\pi\,{\rm Sl}$, and the equation
(\ref{eq:r_spinor}) gives
\begin{equation}
{\sf R}'(2\pi)=\cos\frac{3/2 (2\pi  {\rm Sl})}{2}+\kk\sin\frac{ 3/2(2\pi {\rm Sl})}{2}=\kk^{3 {\rm Sl}}.
\end{equation}
Comparison to $\kk^{\nu^\ast}$ yields the index $\nu^\ast=3
{\rm Sl}$. The factor of $3$ elegantly removes the fractional
quantisation of the self-linking number that stems from the three-fold
symmetry of the profile. This result immediately transforms the
expression (\ref{eq:janichast}) to (\ref{eq:sl_law}) and explains the
correlation between the linking of disclinations and the self-linking
number, discussed in Ref.~\cite{copar_rewiring}.

A similar calculation yields $\phi=\psi/2$ and $\nu^\ast={\rm Sl}$ for
disclinations with a $+1/2$ profile.

There is more to disclinations than just rotations around the
tangent. An obvious possibility is rotation around a perpendicular
axis for $\phi=P\pi$, where $P=\pm1$ is the parity that distinguishes
left-handed and right-handed rotation for $\pi$.  The result of such a
rotation is the transition between $+1/2$ and $-1/2$ defect profiles
(Fig.~\ref{fig:figure2}c). The axis of rotation in equation
(\ref{eq:r_spinor}) can point at any angle $\theta$ in the $xy$ plane,
$\vec{a}=(\cos\theta,\sin\theta,0)$, which leads to
\begin{equation}
 {\sf R}'_\perp=P(\ii\cos\theta+\jj\sin\theta)=P\ii \exp(-\kk \theta).
\end{equation}

The rotations can be composed sequentially in the order they are
encountered with increasing parameter $v$. A real disclination loop
can have a profile that rotates around a gradually varying axis that
is neither parallel nor perpendicular to the disclination
tangent. However, the principal reasoning in our derivation is
topological, so the exact geometric realisation of the rotations is
not important. For example, the location of a transition between
profiles can be pinpointed to the place where the circle's normal is
exactly perpendicular to the disclination's tangent. Whether the
transition between $-1/2$ and $+1/2$ disclination profiles happens
along a very short segment of the disclination, or takes the entire
disclination length to complete, has no effect on the calculated
index, as the director field in the second scenario can be smoothly
combed to the first one without changing the topology. Consequently,
rotations around the $z$ axis (${\sf R}'_z$) and rotations for $\pi$
around a perpendicular axis (${\sf R}'_\perp$) are enough to describe
all director profile variations:
\begin{align}
  {\sf R}'_z&=\cos\frac{\phi}{2}+\kk\sin\frac{\phi}{2}=\exp(\kk \tfrac{\phi}{2}),\\
  {\sf R}'_{\perp}&=P(\ii\cos\theta+\jj\sin\theta)=P\ii \exp(-\kk \theta).
\end{align}
For the rotations ${\sf R}'_z$, the angle $\phi$ is the local rotation angle
of the director, which relates to the \emph{physical} rotation angle $\psi$
of the profile: $\phi=(3/2)\psi$ for a $-1/2$ profile and
$\phi=(1/2)\psi$ for a $+1/2$ profile.

Because of the presence of the unit quaternion $\ii$ in the
perpendicular rotation, it does not commute with $z$-axis rotations.
\begin{equation}
  {\sf R}'_z[\phi]{\sf R}'_\perp[\theta]= {\sf R}'_\perp[\theta]{\sf R}'_z[-\phi]
\end{equation}
The rotation of $+1/2$ has the opposite effect than the rotation of
$-1/2$ profile. The product of all local rotations must satisfy the
continuity condition (\ref{eq:quatern_cycle}). As a perpendicular
rotation introduces a factor $\ii$, they must always appear in pairs
-- if the disclination turns from $-1/2$ to $+1/2$ it must turn back
into $-1/2$. The offset in the axis orientations adds an extra amount
of torsion,
\begin{align}
  {\sf R}'_\perp[P_1,\theta_1]{\sf R}'_\perp[P_2,\theta_2]&=-P_1P_2\exp(\kk(\theta_1-\theta_2)).
\end{align}
If two sequential rotations about the same perpendicular axis are
performed with the same parity (twice by $\pi$ or twice by $-\pi$),
they yield
\begin{equation}
  ({\sf R}'_\perp)^2=-1,
  \label{eq:switchparity}
\end{equation}
which effectively changes the index $\nu^\ast$ by $2$ and switches the
topological charge between even and odd. Opposite parities naturally
cancel out.

As an example, consider a disclination loop with a reference profile
of $-1/2$ at $v=0$, that rotates by $\psi_1$ around its tangent,
flips to a $+1/2$ profile around an axis at an angle $\theta_1$ and with
parity $P_1$, then rotates by $\psi_2$ and flips back to $-1/2$
around an axis at $\theta_2$ with parity $P_2$. The total rotation of
the profile along the entire loop becomes
\begin{align}
  {\sf R}'=&(-P_1 P_2)\ii\exp(-\kk \theta_2)\exp(\kk \tfrac{\psi_2}{4})\ii\exp(-\kk\theta_1)\exp(\kk \tfrac{3 \psi_1}{4})\notag\\
  =&-P_1P_2\exp\left[\kk\left(\frac{3}{4}\psi_1-\frac{1}{4}\psi_2+\theta_2-\theta_1\right)\right].
\end{align}
The exponent must be a multiple of $\pi/2$ to satisfy the continuity
equation, which puts a restriction on the angles $\phi_{1,2}$ and
$\theta_{1,2}$. This generalises the third-integer quantization of the
self-linking number. When this condition is met, the above expression
can be compared to $\kk^{\nu^\ast}$ to extract the index $\nu^\ast$.

Note that the self-linking rotations of $+1/2$ and $-1/2$ parts have
opposite-signed contributions to the index $\nu^\ast$. Taking a
different part of the loop as a reference cyclically permutes the
order of the transformations. For example, if a $+1/2$ part serves as
a reference, the entire exponent changes sign.  This does not affect
the ${\sf R}'=\pm 1$ rotations ($\nu^\ast\in \{0,2\}$), but exchanges
the signs of ${\sf R}'=\pm \kk$ for linked disclinations. The same
effect is observed if the parametrization of any of the disclination
loops is reversed, which also reverses the linking numbers in equation
(\ref{eq:janichast}). This ambiguity stems from the requirement of the
classification with homotopy groups to have a fixed reference -- a
base point \cite{mermin}. The choice of reference is very important
once more than one loop is present in the system.

Most of the disclination loops can be analysed directly by the
described formalism. Wherever the normal of the circle traced by the
director on the unit sphere is not perpendicular to the disclination
tangent, it can be simply regarded as being a $-1/2$ or a $+1/2$
profile and the director can be unambiguously combed to the
perpendicular plane. The decomposition thus takes advantage of easily
recognisable physical features of the disclination profiles and can be
counted by hand if provided with a properly visualised director field.
In singular cases, such as a twist disclination line, which is exactly
half-way between $+1/2$ and $-1/2$ profiles, appropriate quaternion
expressions can be quickly retrieved using a procedure similar to the
one described above.

\section{Nematic surface textures}

\begin{figure}
  \centering
  \includegraphics[width=\textwidth]{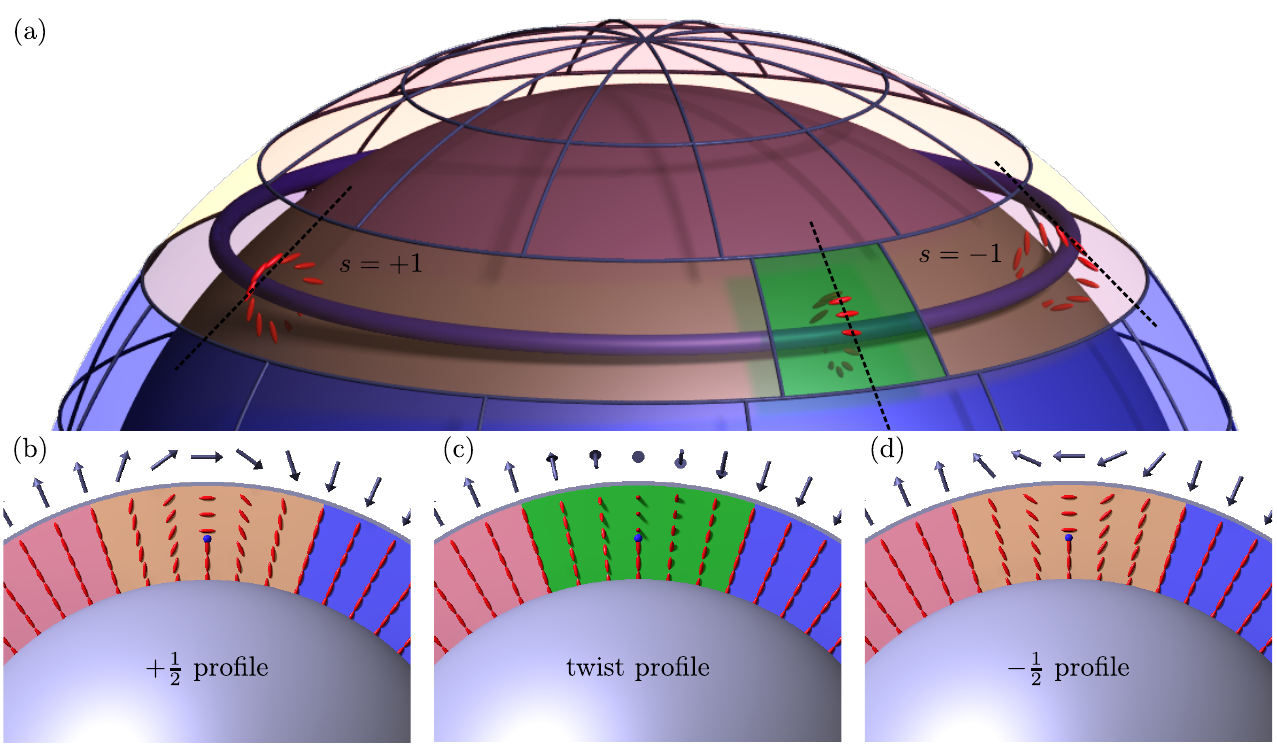}
  \caption{
    \label{fig:figure3}
    The correspondence between subsurface disclination loops and
    geometric features of nematic textures on a sphere-like surface
    (a). The patches of perpendicular director (red for outgoing and blue for ingoing field)
    are segmented by boundaries (orange) with the director parallel to the surface.
    Boundary signature $s=\pm 1$ corresponds to $+1/2$ (b) or $-1/2$ (d) profiles of the
    subsurface disclination loops, while the grains -- places where the director
    is parallel to the boundary -- correspond to the twist profile, which
    is an intermediate state in transition between the other two elementary
    disclination profiles.
  }
\end{figure}

To present above statements from a different perspective, we must
first shortly review the theory behind nematic point defects. The
topological charge of a set of defects contained inside a surface is
determined by the second homotopy group class of the nematic texture
on said surface. A convenient way of calculating the topological
charge is introduced in reference \cite{copar_decomposition}. It
states that a surface texture can be split into patches (2-dimensional
objects) where the director is perpendicular to the enclosing surface,
separated by boundaries (1-dimensional objects) where the director is
parallel to the surface, but perpendicular to the boundary itself, and
grains (isolated points on the boundaries) where the director is also
parallel to the boundary of which it is part. The topological charge
of the texture can then be calculated by counting the contributions of
boundaries $t=\pm 1$ and grains $g=\frac12$.
\begin{equation}
  q=1+\sum_i t_i+\sum_j g_j
\label{eq:texturefull}
\end{equation}
The first term is a half of the Euler characteristic of the enclosing
surface and is related to the fact that a radial hedgehog has a
topological charge of $q=1$.

There is a striking similarity between the classification of
disclination loops and the classification of nematic textures. Both
theories observe the behaviour of the director in a local frame,
aligned to the surface or the curve in question. This splits the
global contribution of the shape's topology from the local geometric
contributions. For disclination loops, the global part is the ${\sf
  R}_0=-1$ factor that describes the nonzero topological charge of a
simple Saturn ring defect, while for surface textures, it is the $+1$
contribution of the Euler characteristic that describes the nonzero
topological charge of a radial hedgehog. Both approaches also assume
the texture or the director profile can be smoothly transformed, so
that all the changes happen in narrow transition zones.

The Figure \ref{fig:figure3} shows that every surface texture can be
alternatively interpreted as a homeotropic core with added
disclination loops below the surface of interest at the places where
the boundaries are, and with disclination profile transitions where
the grains are. The condition that the grains come in pairs and change
the topological charge by $1$ if they have the same parity,
corresponds exactly to the same rule for the profile transitions. The
topological charge of the texture can therefore also be calculated by
adding the core $+1$ defect (the Euler characteristic) to the charges
contributed by J\"anich's indices for all the virtual subsurface
disclination loops below the boundaries,
\begin{equation}
  q=1+n+\sum_j g_j\mod 2,
\label{eq:texturemod}
\end{equation}
for $n$ boundaries and grains $g_j=\pm \frac{1}{2}$. One big
difference between equations (\ref{eq:texturefull}) and
(\ref{eq:texturemod}) is that the result is only given by modulo $2$
in the latter. This is a natural consequence of dealing with line
defects instead of point defects (see Ref.~\cite{randy_rmp}). However,
the narrow subset of disclinations that can act as subsurface
representations of boundaries, are restricted by the surface of the
homeotropic core, so there must be no linking and self-linking.

The correspondence between the textures and disclination loops is not
necessarily only theoretical. If the surface on which we are observing
the texture lies just above the disclinations, the geometric features
on the surface texture will reflect the underlying disclination
geometry.  From this, another strong conclusion can be derived. The
defect rank, a geometric measure for classifications of point charges,
introduced in Ref.~\cite{copar_decomposition}, is only valid if there
are no grains on the surface. By the correspondence principle, we can
state that if there are no disclination transitions, no linking and no
self-linking in the system, the defect rank can be used and the usual
intuition of assigning proper integer charges to both disclination
loops and point defects is valid. As soon as the system becomes
entangled or there are different disclination profiles present in the
system, there are possible discrepancies in the full integer
conservation of the topological charge and only the even and odd
conservation law holds.

\section{Conclusion}

The topological properties of nematics are an interesting topic,
crucial for interpretation and design of experiments. Although the
basic, strictly topological formalism has been known for a long time,
easy application of the theory to the real-world examples requires
geometrization of the basic topological concepts. By considering local
director behavior, we can account for restrictions enforced by the
energy minimisation and thus make a fine-grained classification using
not only the homotopy theory but also additional geometric
information.

We have decomposed a general disclination loop into a sequence of
local rotations that must add up continuously to a rotation that
satisfies the director continuity in a closed disclination
loop. Quaternions are an analytically efficient way of enumerating the
local behavior of the director profile. The use of quaternions further
reinforces the close relationship between the uniaxial and biaxial
nematics, as the unified handling of the entire cross section removes
the differences between both phases.

As an extension of the theory of $-1/2$ defects, the introduced
formalism describes a wider range of disclination networks. Notable
examples of systems that exhibit transitions between different
director profiles are confined cholesteric phases
\cite{fukuda_blue_prl,fukuda_blue_ring}, optically induced defects
\cite{tine_opt} and transient defects that are created with quenching
from the isotropic phase \cite{coarsening}. Studies of random blue
phases \cite{dennistonblue} and entangled structures in random or
structured pores \cite{araki_nmat_mem} could also benefit from
improved classification of disclination profile variations.

The formalism we presented also shows a stunning connection with the
texture decomposition, suggesting that the approaches of
geometrisation for different kinds of defects can be unified, leading
toward a consistent set of tools for enumeration of defects.

\section{Acknowledgments}
This research was funded by Slovenian Research Agency under Contracts
No. P1-0099 and No. J1-2335, NAMASTE Center of Excellence, and
HIERARCHY FP7 network 215851-2. S.~\v{C}. was supported, in part, by
NSF Grant DMR05-47230. Both authors also acknowledge partial support
by NSF Grant PHY11-25915 under the 2012 KITP miniprogram ``Knotted
Fields''.

%\bibliography{copar_zumer_quaternions_nematics_PP_01}

\end{document}